\documentclass{article}
\usepackage{arxiv}
\usepackage[utf8]{inputenc} 
\usepackage[T1]{fontenc}    
\usepackage[colorlinks,
            linkcolor=blue,
            anchorcolor=blue,
            citecolor=green]{hyperref}
\usepackage{url}            
\usepackage{booktabs}       
\usepackage{amsfonts}       
\usepackage{nicefrac}       
\usepackage{microtype}      
\usepackage{lipsum}		
\usepackage{graphicx, subcaption}
\usepackage{doi}
\usepackage{gensymb}
\usepackage{algorithm}
\usepackage{algpseudocode}
\usepackage{amsmath}
\usepackage{color,xcolor,soul}
\usepackage{siunitx}
\usepackage{enumitem}
\usepackage{natbib}
\usepackage{comment}
\setcitestyle{authoryear,round}
\usepackage{multirow}

\makeatletter
\newcommand{\customlabel}[2]{%
   \protected@write \@auxout {}{\string \newlabel {#1}{{#2}{\thepage}{#2}{#1}{}} }%
   \hypertarget{#1}{}
}
\makeatother

\title{Towards an end-to-end artificial intelligence driven global weather forecasting system}

\date{} 					

\author{Kun Chen $^{1,2}$ ~~
	    Lei Bai $^{2,}$\thanks{Corresponding author: baisanshi@gmail.com, eetchen@fudan.edu.cn.}  ~~
        Fenghua Ling $^{2,3}$ ~~
        Peng Ye $^{1,2}$ ~~
        Tao Chen $^{1,*}$ ~~
        Hang Fan $^{2,3}$ ~~
        Hao Chen $^{2}$ ~~ \\ \\
        \textbf{Yi Xiao} $^{2}$ ~~
        \textbf{Kang Chen} $^{2}$ ~~
        \textbf{Tao Han} $^{2}$ ~~
        \textbf{Jing-Jia Luo} $^{3}$ ~~
	      \textbf{Wanli Ouyang} $^{2}$ \\ \\
        $^{1}$ Fudan University ~~
        $^{2}$ Shanghai AI Laboratory ~~
        $^{3}$ Nanjing University of Information Science and Technology
}

\renewcommand{\shorttitle}{\textit{arXiv} Template}

\hypersetup{
pdftitle={A template for the arxiv style},
pdfsubject={q-bio.NC, q-bio.QM},
pdfauthor={David S.~Hippocampus, Elias D.~Striatum},
pdfkeywords={First keyword, Second keyword, More},
}

\begin{document}
\maketitle

\begin{abstract}
	The weather forecasting system is important for science and society, and significant achievements have been made in applying artificial intelligence (AI) to medium-range weather forecasting. However, existing AI-based weather forecasting models rely on analysis or reanalysis products from traditional numerical weather prediction (NWP) systems as initial conditions for making predictions. The initial states are typically generated by traditional data assimilation components, which are computationally expensive and time-consuming. Here, by cyclic training to model the steady-state background error covariance and introducing the confidence matrix to characterize the quality of observations, we present an AI-based data assimilation model, i.e., Adas, for global weather variables. Further, we combine Adas with the advanced AI-based forecasting model (i.e., FengWu) to construct an end-to-end AI-based global weather forecasting system: FengWu-Adas. We demonstrate that Adas can assimilate global conventional observations to produce high-quality analysis, enabling the system to operate stably for long term. Moreover, the system can generate accurate end-to-end weather forecasts with comparable skill to those of the IFS, demonstrating the promising potential of data-driven approaches.
\end{abstract}


\section{Introduction}
Driven by the advancements and maturity of AI, particularly deep learning techniques, scientific intelligence has been rapidly evolving with the aim of leveraging AI to promote scientific research and discovery. Within the field of atmospheric science, AI has achieved remarkable achievements in various areas, such as post-processing and bias correction~\citep{yang2023improving,mouatadid2023adaptive, agrawal2023machine, singh2022short}, downscaling~\citep{harder2022generating, chen2022rainnet}, precipitation nowcasting~\citep{ravuri2021skilful, gao2022earthformer, zhang2023skilful}, climate forecasting~\citep{ling2022multi}, and medium-range weather forecasting~\citep{pathak2022fourcastnet, bi2023accurate, lam2023learning, chen2023fengwu, chen2023fuxi}. Some AI-based models have demonstrated highly competitive forecasts compared to the deterministic forecasts of the state-of-the-art NWP system, i.e., the Integrated Forecasting System (IFS) from the European Centre for Medium-Range Weather Forecasts (ECMWF). These models are usually trained on reanalysis dataset (e.g., ERA5~\citep{hersbach2020era5} from ECMWF) and allow much lower computational costs and easier deployment for operational forecasting. Despite drawbacks such as forecast smoothness, bias drift and lack of physical consistency~\citep{ben2023rise, bonavita2024some}, AI approaches have shown the immense potential of data-driven modeling in weather prediction, offering a new paradigm for meteorological forecasting.

\begin{figure}
    \centering
    \includegraphics[width=\linewidth]{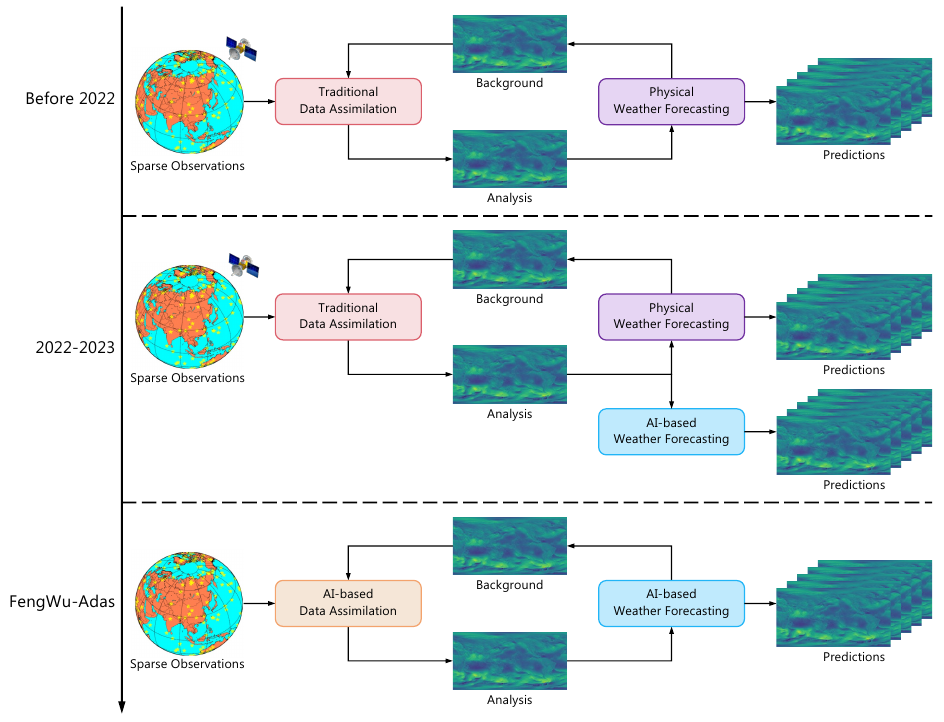}
    \caption{\textbf{The progression of global weather forecasting system.} Traditional NWP systems consist of physical weather forecasting model and data assimilation. The breakthrough of AI-based medium-range weather forecasting models occurred in 2022-2023 with highly competitive performance in terms of accuracy, but they still rely on the NWP systems for making predictions. Our work is dedicated to exploring the possibility of an end-to-end global weather forecasting system which is driven purely by AI.}
    \label{fig:timeline}
\end{figure}

However, the AI-based weather forecasting models still require analysis products generated through the process of \textit{data assimilation} in the traditional NWP system for making predictions (Figure~\ref{fig:timeline}). Specifically, data assimilation aims to obtain the best estimate of the true state of the Earth system (known as the \textit{analysis}) and provide an accurate initial state for weather prediction, thus improving the forecast performance. In the operational weather forecasting system, data assimilation is a critical component that ensures the long-term stable operation of the system by periodically correcting the forecast. Beyond the predictability limit, which is believed to be about 15 days, the forecast results without data assimilation will become completely unreliable. This process is heavily dependent on observations, as they provide crucial information that closely represents the true state of the atmosphere. For example, the earliest initial conditions were obtained by interpolating observations onto the grid points of the grid space~\citep{richardson1922weather}. Modern data assimilation techniques are usually achieved by integrating observations with short-range forecasts (i.e., the \textit{background}), primarily including two categories: Kalman filters and variational methods~\citep{rabier2003variational}. The main motivation for introducing the background, or the first guess, lies in the incomplete nature of observations. The sparsity and irregularity of observations make it difficult to obtain a global estimate, resulting in low-quality interpolation analysis. Besides, not all instruments can observe geophysical variables directly.

Currently, AI methods have been applied primarily to specific processes within data assimilation to address the limitations of traditional data assimilation techniques~\citep{buizza2022data, cheng2023machine}. For example, implicit neural representations (INRs)~\citep{li2023latent} and various autoencoders (AEs)~\citep{peyron2021latent, melinc2023neural, amendola2021data} have been employed as reduced-order models (ROMs) to address high-dimensional challenges. These methods replace classical linear models such as proper orthogonal decomposition (POD)~\citep{cheng2023generalised, pawar2022equation} in latent assimilation, providing an efficient framework for the representation and reconstruction of geophysical variables in latent space. Neural networks have also been utilized to derive tangent-linear and adjoint models in 4D variational (4DVar) data assimilation~\citep{hatfield2021building}, provide localization functions for ensemble Kalman filters (EnKF)~\citep{wang2023convolutional} and estimate error covariance matrix~\citep{cheng2022observation, penny2022integrating}. Moreover, there exist strong mathematical similarities between machine learning and data assimilation, particularly the variational data assimilation~\citep{geer2021learning}, which enables the optimization of the cost function in variational data assimilation using auto-differentiation~\citep{melinc2023neural, xiao2023fengwu}. However, these studies have not involved fundamentally altering the algorithms of data assimilation.
While traditional data assimilation algorithms are underpinned by rigorous mathematical theory and physical models, they inevitably rely on certain assumptions as prerequisites. In practical operational applications, they often face constraints due to high computational costs, which requires various approximations for solution~\citep{carrassi2018data}. In situations where supervised information is available, neural networks are expected to automatically capture the correlations among data during the assimilation and provide a new alternative algorithm~\citep{fablet2021learning}. Efforts have been made in idealized data assimilation experiments~\citep{rozet2023score,huang2024diffda,chen2024fnp}, and further research focuses on data-driven end-to-end weather forecasting systems, broadly on systems that can initialize forecasts without needing traditional physically-based analysis~\citep{vaughan2024aardvark,xu2024fuxi,mcnally2024data}. Some follow the paradigm of traditional NWP systems and specially design data assimilation models to work together with AI-based weather forecasting models in a forecast-assimilation cycle to form a system~\citep{xu2024fuxi}. Others implicitly integrate the process of data assimilation and weather forecasting, using a single model to make predictions directly from observations~\citep{vaughan2024aardvark,mcnally2024data}.

In this study, we present Adas, a novel data assimilation model for global weather variables, which can assimilate sparse conventional observations with different qualities and provide initial conditions for weather forecasting. The key of data assimilation lies in adjusting the weights of the background and observations based on error covariance, and traditional data assimilation techniques primarily derive optimal analysis under the Bayesian framework. In our method, we model the steady-state (i.e., optimal dynamic balance between forecasting error growth and assimilation error reduction) background error covariance by cyclic training and introduce the confidence matrix to characterize the availability and quality of observations during the data assimilation. With the guidance of confidence matrix, Adas employs the gated convolution module to handle sparse observations and the gated cross-attention module for capturing the interactions between the background and observations efficiently. The prediction of the advanced AI-based weather forecasting model, FengWu~\citep{chen2023fengwu}, is used to generate the background for data assimilation. In return, the analysis of Adas is then used as the initial state of FengWu for making predictions at the next time step, thus forming an end-to-end AI-based global weather forecasting system: FengWu-Adas. We demonstrate that our system can operate stably for long term, and our method is robust and flexible to be combined with any state-of-the-art weather forecasting model without retraining or fine-tuning. The ideal experiments have demonstrated the superior properties of FengWu-Adas in producing accurate analysis. Further, we apply FengWu-Adas to real-world scenarios, conducting a comprehensive and detailed evaluation to the system and making comparison with the IFS. The results show that FengWu-Adas can generate high-quality analysis and accurate end-to-end global weather forecasts although it only uses conventional observations, demonstrating the promising potential of data-driven approaches.

\section*{Results}\label{Results}
\subsection*{Principle of FengWu-Adas}\label{Principle of FengWu-Adas}

\begin{figure}[h]
    \centering
    \includegraphics[width=\textwidth]{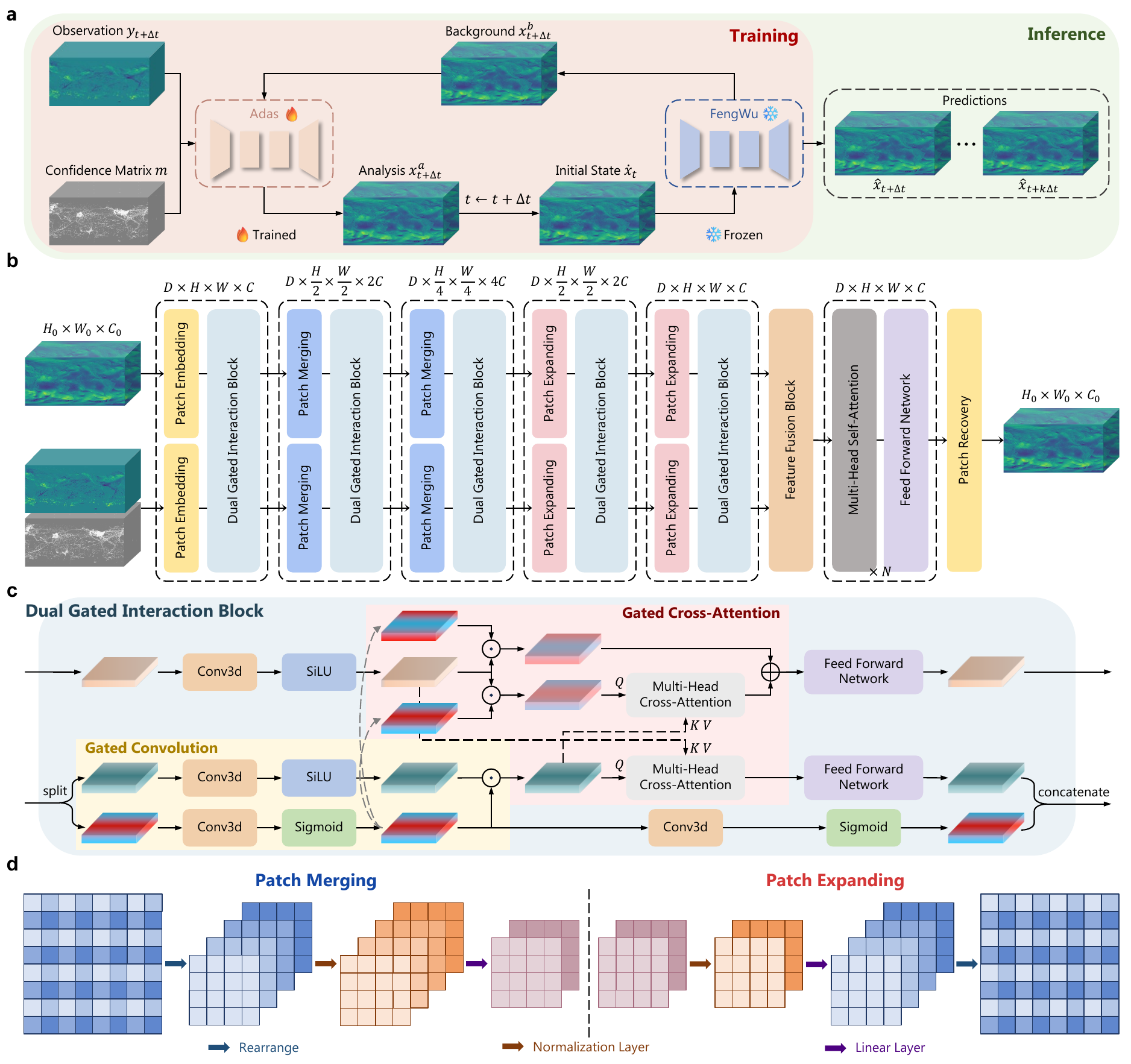}
    \caption{\textbf{The framework of FengWu-Adas and network architecture of Adas.} \textbf{a}, The framework and process of FengWu-Adas. The prediction of FengWu provides the background for data assimilation and the analysis of Adas serves as the initial state for weather forecasting. The parameters of FengWu are frozen during training and not further fine-tuned or involved in joint optimization. \textbf{b}, Overview of Adas's network structure. The dual encoder composed of gated interaction blocks extract multi-scale features of the background and sparse observations separately and capture the interactions between them, and then the features are fused and decoded. \textbf{c}, Details of dual gated interaction block. The gated convolution and gated cross-attention modules are designed for sparse observations, which are both guided by the confidence matrix. \textbf{d}, Simple schematic diagram of patch merging and patch expanding. The patch merging performs the rearrangement operation and linear layer to achieve down-sampling and the patch expanding performs the opposite operation to achieve up-sampling.}
    \label{fig:principle} 
\end{figure}

The framework of FengWu-Adas is shown schematically in Figure~\ref{fig:principle}a. For the given initial state denoted as $\dot{x}_t$ at time $t$, FengWu can make multi-step predictions $\hat{x}_{t+\Delta t}$, $\hat{x}_{t+2\Delta t}$, \dots, $\hat{x}_{t+k\Delta t}$ at time $t+\Delta t$, $t+2\Delta t$, \dots, $t+k\Delta t$ in an auto-regressive manner, where $k$ denotes the time index and $\Delta t=6h$ is the temporal spacing of single-step prediction. Considering that the accuracy of the background will directly affect the quality of the analysis, the single-step prediction of FengWu is used as the background $x_{t+\Delta t}^b$ for data assimilation. After obtaining the observations $y_{t+\Delta t}$ at the corresponding time, confidence levels ranging from 0 to 1 can be assigned to each value in observation field based on the availability and quality of the observation, thereby generating the confidence matrix. The confidence matrix $m$, serving as additional information for the uncertainty of observation error, is sent into Adas along with observations for assimilation. For Adas, the input dataset is the background and observations, while the target dataset is ERA5, and the learning objective is to integrate the background forecast and observations to obtain analysis close to ERA5. Subsequently, the analysis $x_{t+\Delta t}^a$ produced by Adas will be the initial state for the next prediction, i.e., the input of FengWu at next time and so on, in a cyclic manner. Different from the general single-step training with a fixed lead background, this cyclic training makes the model more robust to the background, because the background quality by starting the forecast with analysis at the previous time changes dynamically with the convergence of the model training. By continuously iterating to optimize the implicit modeling of background error covariance, the system can achieve a steady state that balances the error growth of forecasting and the error reduction of assimilation. This approach is not limited to the optimal analysis of single-step assimilation, instead seeking a dynamic equilibrium for the forecasting and assimilation of the entire system, whose superiority has been previously substantiated~\citep{yang2006data,kalnay20074}. As an end-to-end global weather forecasting system, FengWu-Adas can also perform multi-step forecasts through auto-regression based on the analysis.

Adas utilizes a dual encoder composed of gated interaction blocks to extract features of the background and sparse observations separately and capture the interactions between them (Figure~\ref{fig:principle}b). All the inputs with a shape of $H_0\times W_0\times C_0$ are converted into a unified latent shape of $D\times H\times W\times C$ through the patch embedding module so that the network can capture their spatial relationships in the 3D mesh. $H_0$ and $W_0$ represent the number of grid points in the longitudinal and latitudinal directions, respectively, and the channel $C_0$ represents the number of variables. The 3D upper-air variables and 2D surface variables are embedded into the latent space independently and then concatenated together, where the information at each grid point is represented by a $C$-dimensional vector. $H$ and $W$ correspond to the number of grid points in the longitudinal and latitudinal directions after embedding, and $D$ represents the number of levels in the vertical direction. By performing down-sampling and up-sampling through the patch merging and patch expanding module (Figure~\ref{fig:principle}d), the encoder can capture multi-scale meteorological features. The dual gated interaction block extracts features from the background and observations with standard convolution and gated convolution, along with efficient information interactions through the gated cross-attention module (Figure~\ref{fig:principle}c). Both the gated convolution and gated cross-attention are guided by the confidence matrix, which is updated together with the features to represent the availability and quality of observations (details in Methods). After the feature fusion block implemented by a convolutional layer, the features are sent into a series of Transformer~\citep{vaswani2017attention, dosovitskiy2020image} blocks and then recovered to the original size. The convolutional operation can bring inductive bias of locality, and the attention mechanism enables the model to capture long-range dependencies. The attention block utilizes the shifted-window mechanism~\citep{liu2021swin} adjusted according to geographical rules to reduce the computational cost and ensure continuity in the latitudinal direction.

\subsection*{Superiority of FengWu-Adas in ideal experiments}\label{Superiority of FengWu-Adas in ideal experiments}

\begin{figure}
    \centering
    \includegraphics[width=\linewidth]{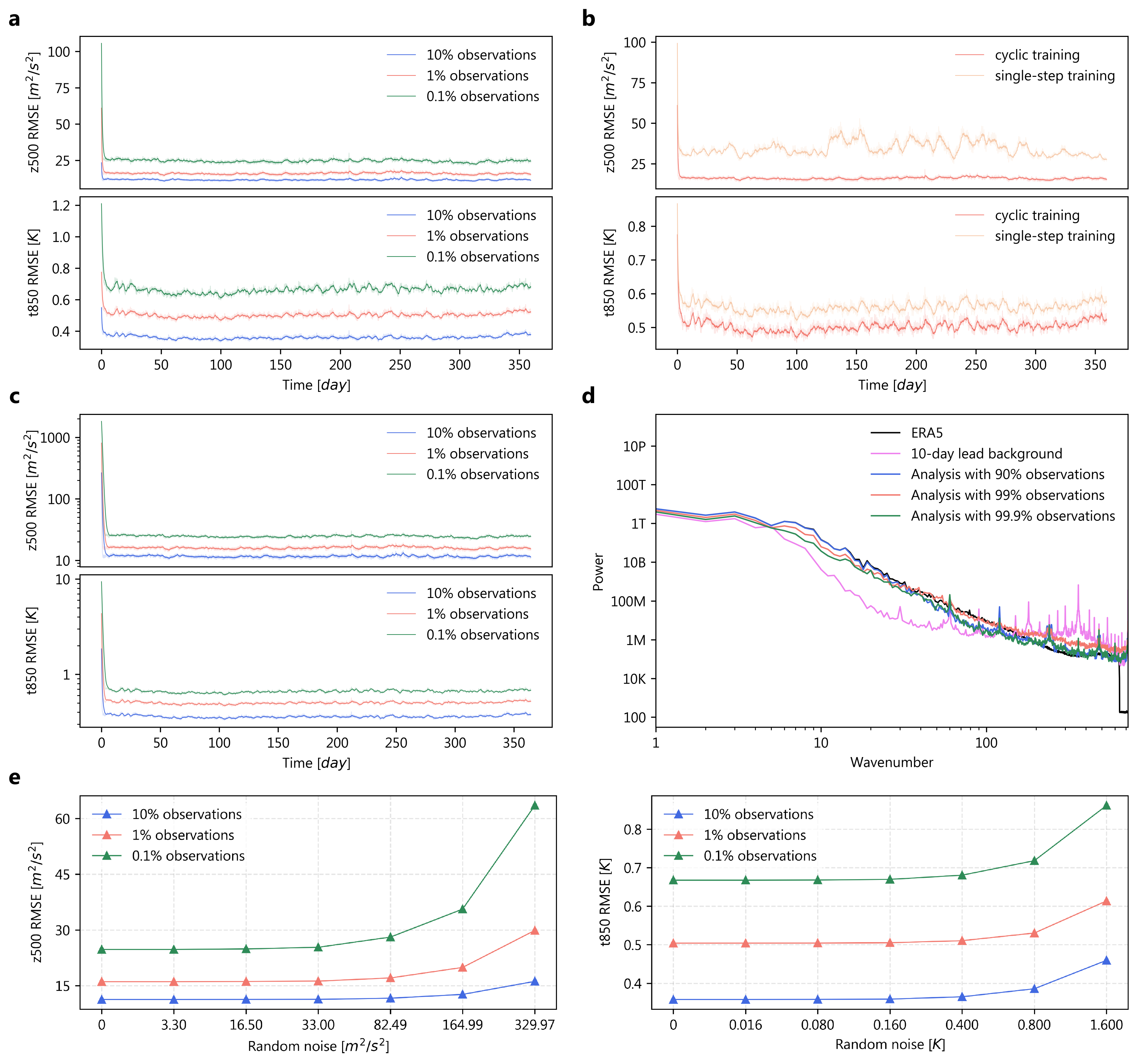}
    \caption{\textbf{The performance and properties of FengWu-Adas in ideal experiments.} The RMSE (\textit{lower is better}) is calculated against ERA5. \textbf{a}, RMSE variations of the analyses for $z500$ and $t850$ variables with different observation ratios in a whole year. The forecast with a 5-day lead time is used as the initial background to start the system. As the system continuously executes the cyclic forecast-assimilation, the RMSE of the analysis decreases rapidly and converges to a lower level. \textbf{b}, The performance differences between cyclic training and single-step training. The cyclic training directly models the background error covariance at steady state, circumventing the issue of error accumulation present in single-step training during cyclic forecast-assimilation experiments and achieving lower steady-state errors. \textbf{c}, The performance when starting the system with completely random Gaussian noise as the initial background. The system can still quickly recover to normal steady-state levels, showing outstanding stability and generalization capability to assimilate observations on arbitrary background. \textbf{d}, The power spectra of the 10-day lead background and analysis for $z500$ variable. After data assimilation, Adas can significantly correct the spectral degradation problem of FengWu forecasts with long lead time at small and medium scales. \textbf{e}, The sensitivity of the system to the noises with varying magnitudes. The horizontal axes correspond to different random Gaussian noise proportions of 0, 0.1\%, 0.5\%, 1\%, 2.5\%, 5\%, and 10\%, respectively. It demonstrates the robustness and anti-interference capability of Adas in the presence of noise.}
    \label{fig:ideal}
\end{figure}

The ideal cyclic forecast-assimilation experiments are first conducted using the ERA5 dataset to verify the performance and characteristics of FengWu-Adas. The experiments proceed with alternating weather forecasting and data assimilation steps: the single-step forecast serves as the background for data assimilation, and the analysis provides the initial state for the next forecasting step. ERA5 is considered as the ground truth, and simulated observations are obtained by sampling it with binary mask, which is then used as the confidence matrix for data assimilation. For simplicity, all variables are sampled independently in a certain proportion. Simulated masks for training are generated randomly at each time step to enhance robustness with respect to the observation locations. During the testing period, however, the masks remain fixed to avoid spurious performance caused by variations in information coverage at different times. The system is initialized with a 5-day lead forecast as the background, and except for producing simulated observations, ERA5 data is invisible throughout the entire experiments. Figure~\ref{fig:ideal}a illustrates the RMSE (defined in Methods) variations of the analyses for $z500$ and $t850$ variables over the year, and the RMSE is calculated against ERA5. The results with different observation ratios are represented by different colors, as marked in the legend. For clarity, the original data is presented with reduced opacity, while solid lines indicate the smoothed values after exponential moving average (EMA). The RMSE of the analyses with all observation ratios rapidly decreases and remains at a low level for the whole year, showing the long-term stability of the system and its ability to generate accurate analysis based on Adas. It should be emphasized that FengWu is trained on ERA5 data~\citep{chen2023fengwu} and is not fine-tuned to our task. As the observation ratio (i.e., the amount of observation information) increases, the speed at which the analysis RMSE decreases becomes faster, and the analysis RMSE in the final steady state is also lower, which is consistent with our general understanding.

We compare the RMSE variations of the analyses under single-step training and cyclic training to demonstrate the effectiveness of the training method in Figure~\ref{fig:ideal}b. The cyclic training directly models the background error covariance at steady state, circumventing the issue of error accumulation present in single-step training during cyclic forecast-assimilation experiments and achieving lower steady-state errors. Moreover, as the model encounters a more diverse range of background samples during training, it exhibits enhanced robustness to backgrounds with varying qualities and results in reduced error fluctuations. With this favorable property, even when starting the system with completely random Gaussian noise as the initial background, it can converge rapidly to the same steady-state level within a few days, and the convergence speed accelerates as the number of observations increases (Figure~\ref{fig:ideal}c). This ensures that the system can correct back to the stable state in a short period of time without affecting subsequent operations even if it encounters a malfunction that leads to complete derailment of the forecast. The generalization capability allows Adas to assimilate observations on arbitrary background without retraining or fine-tuning, so it is flexible to be integrated with any state-of-the-art weather forecasting model as a plug-and-play component.

A major challenge of current AI-based forecasting models is the smoothness of their predictions, which often manifests itself as degraded spectra~\citep{kochkov2024neural,lang2024aifs}. Preserving fine-scale information is also important for data assimilation, allowing for better representation of errors at small and medium scales, which then grow to affect larger scales. Figure~\ref{fig:ideal}d shows the power spectra of the 10-day lead background and analysis for $z500$ variable. After data assimilation, Adas can significantly correct the spectral degradation problem of FengWu forecasts with long lead times at small and medium scales. And as the amount of observation information increases, the spectra of analysis will be closer to that of ERA5. In addition, we examine the sensitivity of the system to the noises with varying magnitudes and the balance problem when assimilating only temperature observations. Figure~\ref{fig:ideal}e shows the changes in RMSE of $z500$ and $t850$ with random Gaussian noise proportions of 0, 0.1\%, 0.5\%, 1\%, 2.5\%, 5\%, and 10\%, respectively. The change in analysis error caused by observational error is much smaller than the observational error itself, demonstrating the robustness and anti-interference capability of Adas in the presence of noise. As the number of observations increases, the model's sensitivity to observational error will decrease further.

\subsection*{Applying FengWu-Adas to real-world scenarios}\label{Applying FengWu-Adas to real-world scenarios}

\begin{figure}
    \centering
    \includegraphics[width=\linewidth]{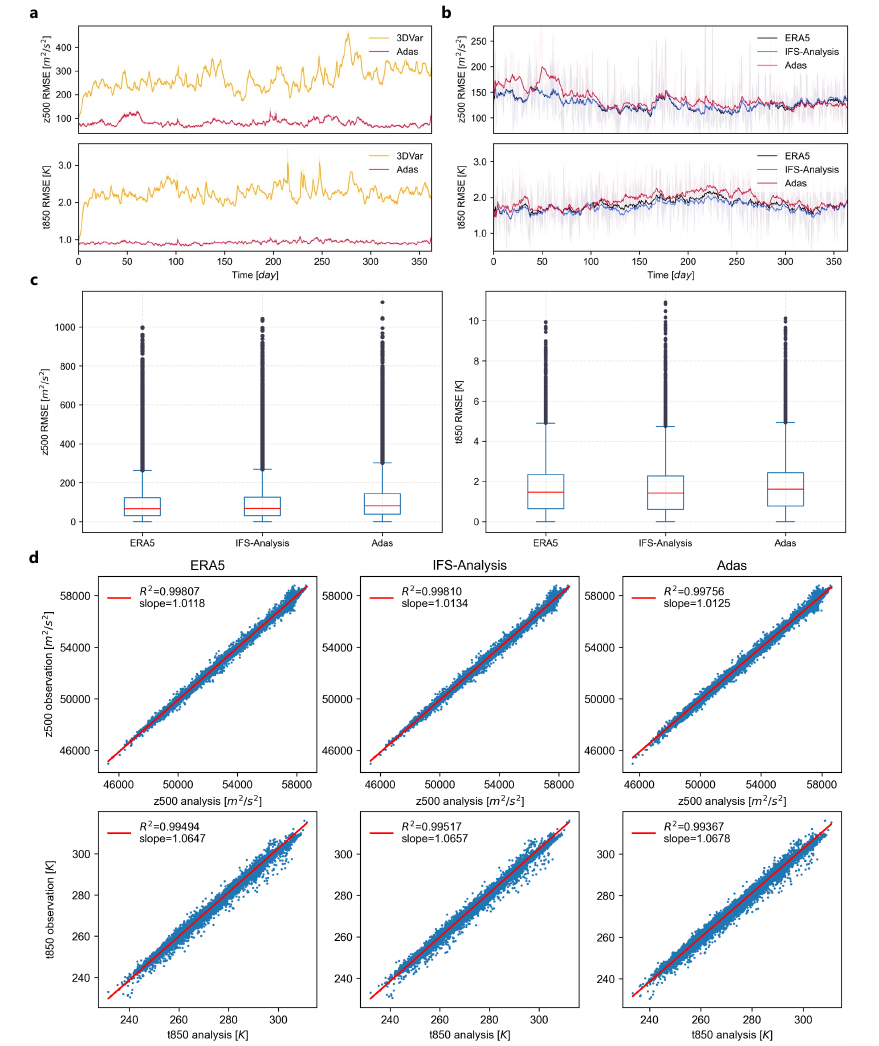}
    \caption{\textbf{The performance of FengWu-Adas to produce analysis with GDAS real observations.} \textbf{a}, RMSE variations of the analyses for $z500$ and $t850$ variables evaluated on ERA5 and the comparison with 3DVar algorithm. By alternating between forecasting and data assimilation, the system can also maintain long-term stability in real-world scenarios. \textbf{b}, RMSE variations of the analyses evaluated on 200 reserved GDAS observation columns and comparison with IFS-Analysis and ERA5. The analysis quality of Adas evaluated at the stations shows a comparable level to IFS-Analysis and ERA5, although Adas only assimilates conventional observations. \textbf{c}, The box plot of the RMSE across the evaluation stations at all times. Whether considering the median, IQR or outliers, Adas shows reasonable results and is not significantly worse than IFS-Analysis and ERA5. \textbf{d}, The scatter plot and corresponding fits between the observation and analysis values across the evaluation stations at all times.}
    \label{fig:gdas_analysis}
\end{figure}

\begin{figure}
    \centering
    \includegraphics[width=\linewidth]{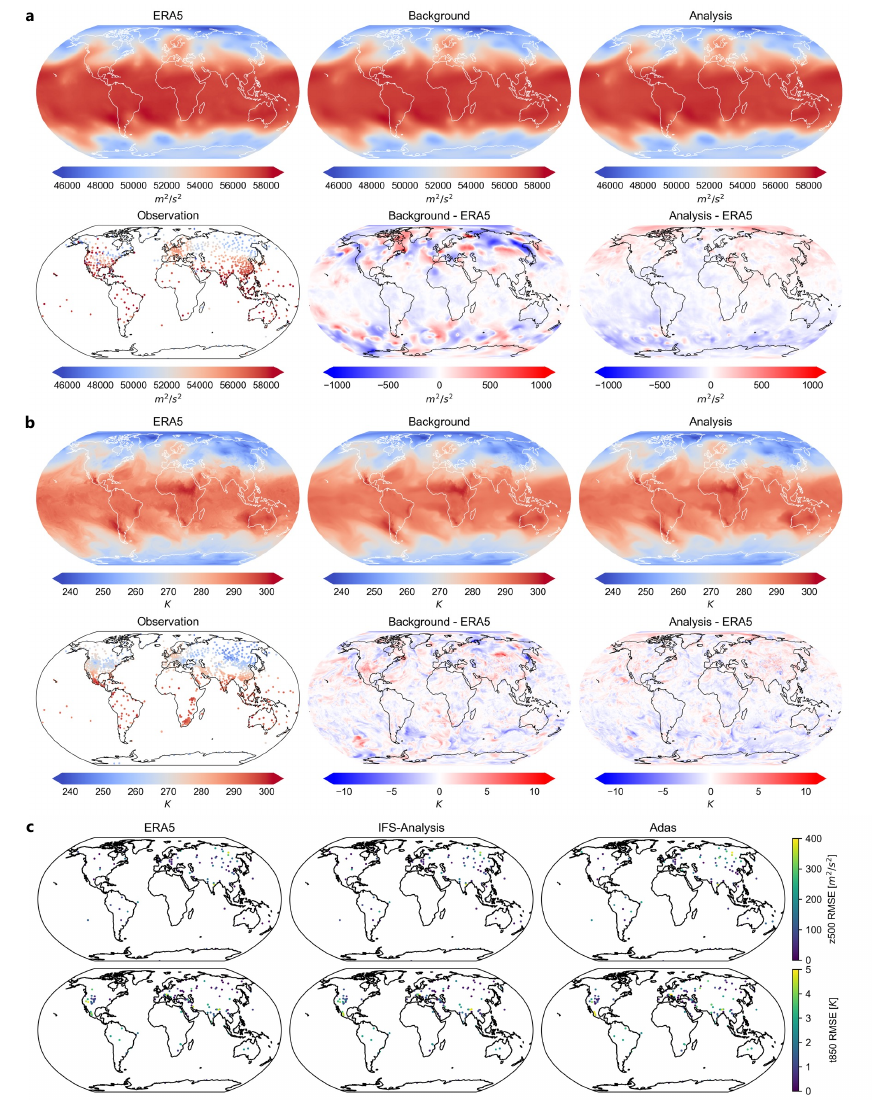}
    \caption{\textbf{Visualization of the fields during data assimilation and the RMSE distribution evaluated on the reserved stations.} \textbf{a}, Visualization of $z500$ variable and corresponding error distributions with ERA5 during data assimilation. The date-time is randomly selected at 2017-01-27 00:00 UTC and the background uses the $96h$ forecast based on ERA5. \textbf{b}, Visualization of $t850$ variable and error distributions during data assimilation. \textbf{c}, Visualization of the RMSE distribution at the reserved stations and comparison with IFS-Analysis and ERA5. The date-time is also selected at 2017-01-27 00:00 UTC, and the analysis of Adas shows similar errors to IFS-Analysis and ERA5.}
    \label{fig:visual_analysis}
\end{figure}

Data assimilation for real observational data is more challenging than simulation experiments and has considerable practical application potential. To validate the performance in real-world scenarios, we apply FengWu-Adas to the assimilation for conventional observations from the Global Data Assimilation System (GDAS). The main data types and information of GDAS observations we use are given in Table~\ref{tab:gdas}. The values of geophysical variables are processed into grid data through nearest neighbor interpolation to generate the observation fields, and the confidence matrix is produced according to the availability and quality of the observations (details in Methods). Figure~\ref{fig:gdas_analysis}a and Figure~\ref{fig:gdas_analysis}b illustrate the RMSE variations of the analyses for $z500$ and $t850$ variables over the year in the cyclic forecast-assimilation experiment, which are calculated against ERA5 and the 200 reserved GDAS observation columns, respectively. When FengWu-Adas is applied to the assimilation for real observational data, the system can still operate stably for long term. Compared with the baseline of 3D variational (3DVar) data assimilation, Adas has lower analysis error and stronger stability. It is worth noting that the 3DVar we reproduced may not be optimal, and it may achieve improved performance with careful tuning. It is encouraging that the analysis quality of Adas evaluated at the stations shows a comparable level to IFS-Analysis and ERA5, although Adas only assimilates conventional observations. Furthermore, we analyze the statistical characteristics of the RMSE across the evaluation stations at all times and draw the box plot in Figure~\ref{fig:gdas_analysis}c. Whether considering the median, interquartile range (IQR) or outliers, Adas shows reasonable results and is not significantly worse than IFS-Analysis and ERA5. Figure~\ref{fig:gdas_analysis}d shows the scatter plot and corresponding fits between the observation and analysis values across the evaluation stations at all times, with the coefficient of determination $R^2$ and the slope of the fitted line. The analysis values of Adas fit the observations well, and its $R^2$ is only slightly lower than that of IFS-Analysis and ERA5. Figure~\ref{fig:visual_analysis}a and Figure~\ref{fig:visual_analysis}b visualize the fields of $z500$ and $t850$ variables and their error distributions with ERA5 during data assimilation, respectively, where the background uses the $96h$ lead forecast based on ERA5 and the date-time is randomly selected at 2017-01-27 00:00 UTC. Based on the information provided by observations, Adas can effectively reduce the background error, and the analysis in areas with dense observations will be more accurate. Figure~\ref{fig:visual_analysis}c visualizes the RMSE distribution evaluated on the reserved stations at the same date-time, they all show similar errors since the learning target of Adas is ERA5.

\begin{table}[h]
    \centering
    \caption{\textbf{Main data information of GDAS observations used in FengWu-Adas system.} GDAS observational data consists of many messages with different types, each of which contains many observation columns with different height levels. The conventional types of observations in GDAS are used for data assimilation and evaluation in our experiments, covering the main geophysical variables such as temperature, humidity and wind speed. The table shows the number of messages and observation columns of each data type at 2017-01-01 00:00 UTC. Upper-air observations are mainly provided by ADPUPA and AIRCFT, and each observation column of ADPUPA usually contains 1 to more than 100 unfixed height levels. Surface observations of land and ocean are mainly provided ADPSFC and SFCSHP, respectively.}
    \label{tab:gdas} 
    \begin{tabular}{cccc} 
        \toprule 
        \multirow{2}*{Type} & Number of & Number of & \multirow{2}*{Description} \\
        & Messages & Obs. Columns & \\
        \midrule 
        \multirow{2}*{ADPUPA} & \multirow{2}*{338} & \multirow{2}*{1403} & Upper-Air (RAOB, PIBAL, RECCO, \\ 
        & & & DROPS) Reports \\
        \multirow{2}*{AIRCFT} & \multirow{2}*{21} & \multirow{2}*{2255} & AIREP/PIREP, AMDAR(ASDAR/ACARS), \\
        & & & E-ADAS(AMDAR BUFR) ACFT \\
        ADPSFC & 1117 & 122093 & Surface Land (SYNOP, METAR) Reports \\
        \multirow{2}*{SFCSHP} & \multirow{2}*{150} & \multirow{2}*{20268} & Surface Marine (SHIP, BUOY, \\ 
        & & & C-MAN Platform) Reports \\
        \bottomrule 
    \end{tabular}
\end{table}

\begin{figure}
    \centering
    \includegraphics[width=\linewidth]{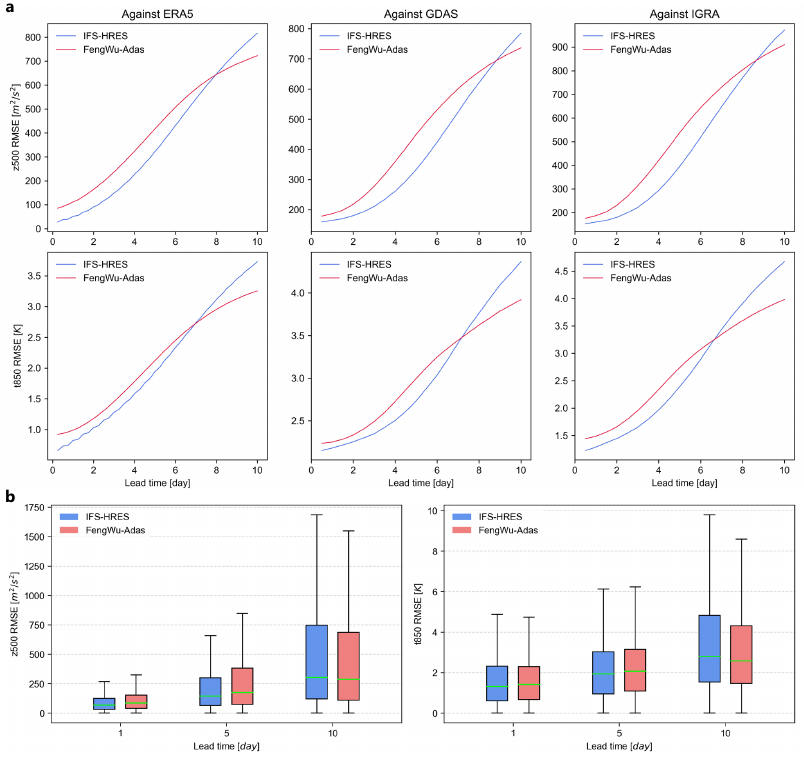}
    \caption{\textbf{The performance of FengWu-Adas to produce end-to-end global weather forecasts and the comparison with IFS-HRES.} \textbf{a}, Annual average RMSE skill for $z500$ and $t850$ variables over 10 days evaluated on ERA5, GDAS and IGRA datasets. All evaluations show the same trend, with FengWu-Adas having higher prediction RMSE than IFS-HRES in the early stage and lower than IFS-HRES starting around 7-8 days. FengWu-Adas shows accurate end-to-end global weather forecasts, demonstrating the promising potential of data-driven approaches. \textbf{b,} The box plot of the prediction RMSE across the GDAS evaluation stations with different lead times. Similar to the results of average RMSE, FengWu-Adas performs slightly worse than IFS-HRES at shorter lead times, but performs better at longer lead times with lower medians and IQRs.}
    \label{fig:gdas_forecast}
\end{figure}

\begin{figure}
    \centering
    \includegraphics[width=\linewidth]{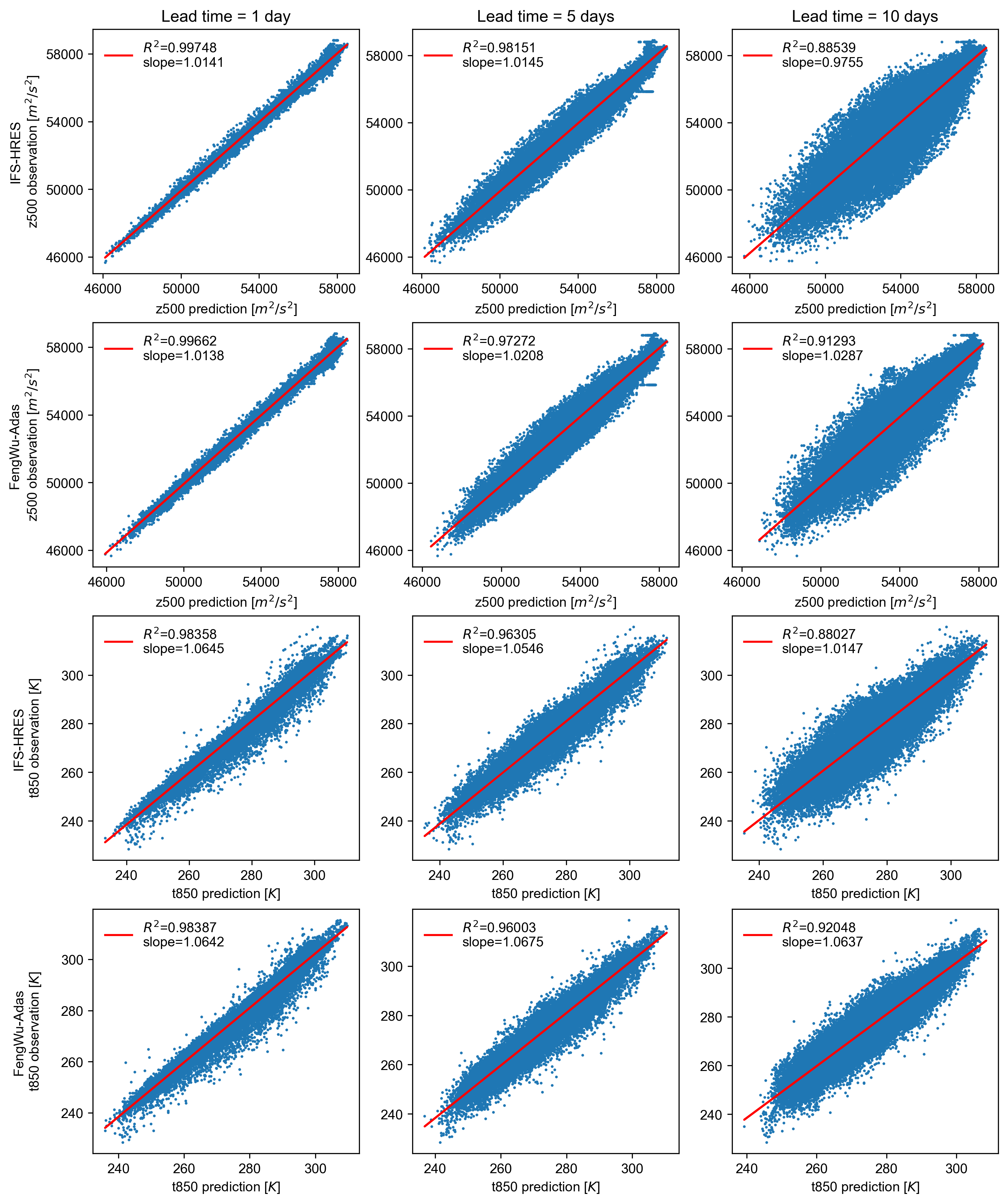}
    \caption{\textbf{The scatter plot and corresponding fits between the observation and prediction values across the evaluation stations.} The first two rows and last two rows correspond to the $z500$ and $t850$ variables, respectively. The columns have different lead times of 1, 5 and 10 days, and the coefficient of determination $R^2$ and the slope of the fitted line are marked in the subplot. The prediction of FengWu-Adas is more consistent with the observations at longer lead times and has a higher $R^2$.}
    \label{fig:forecast_regression}
\end{figure}

\begin{figure}
    \centering
    \includegraphics[width=\linewidth]{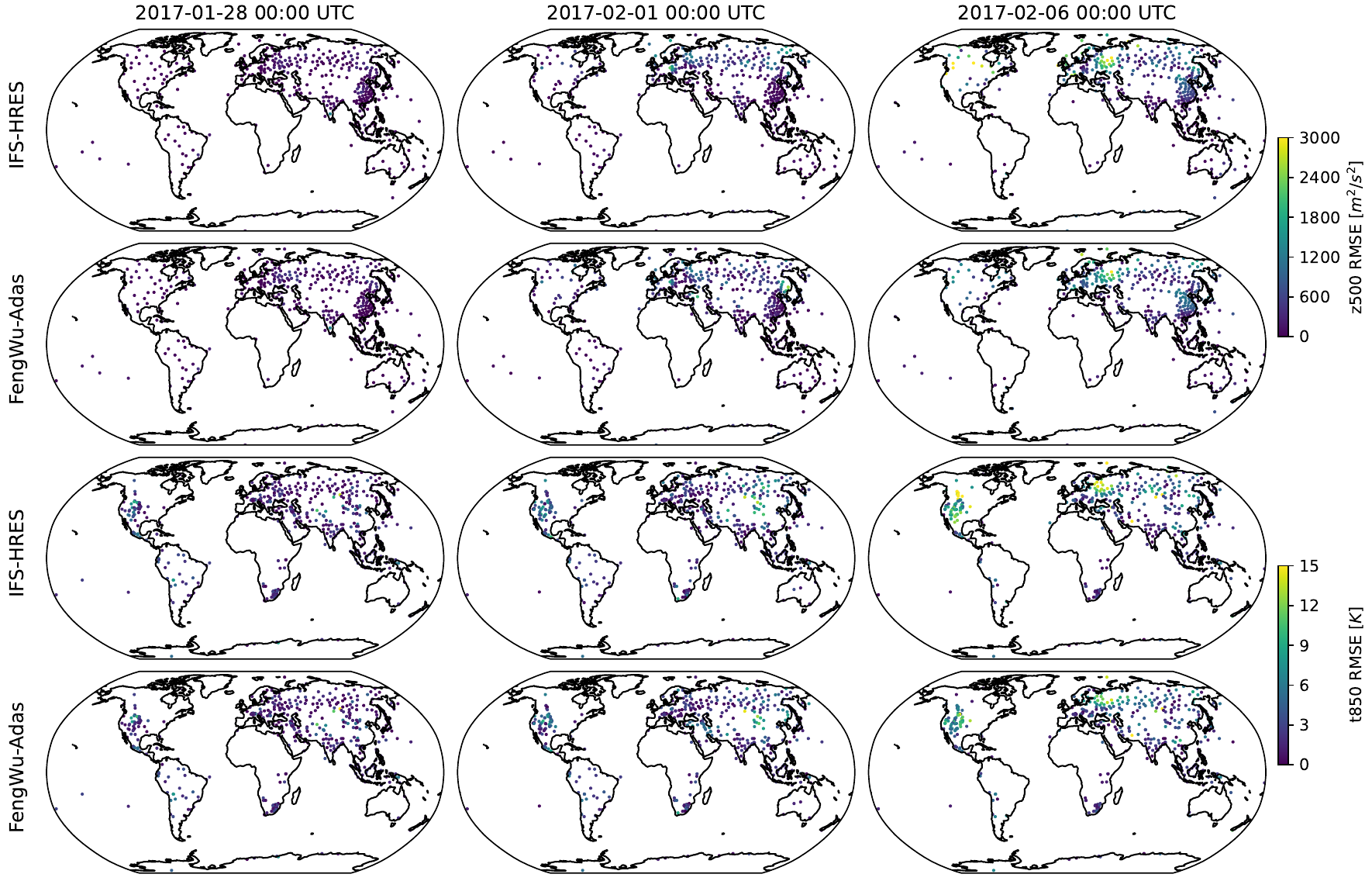}
    \caption{\textbf{Visualization of the RMSE comparison of FengWu-Adas and IFS-HRES forecasts for $z500$ and $t850$ variables at the GDAS stations.} The initialization date-time is set at 2017-01-27 00:00 UTC, which is consistent with Figure~\ref{fig:visual_analysis}, and the RMSE is calculated against GDAS observations. The columns have different lead times of 1, 5, and 10 days, and the first two rows and last two rows correspond to the $z500$ and $t850$ variables, respectively. In the forecast results with 10-day lead time, IFS-HRES has a large deviation in North America, while FengWu-Adas has a better forecast performance in this region.}
    \label{fig:pred_rmse_station}
\end{figure}

To comprehensively evaluate the end-to-end forecasting skill of FengWu-Adas, we test its performance on ERA5, GDAS, and the Integrated Global Radiosonde Archive (IGRA) datasets and compare it with the IFS-HRES. IGRA covers radiosonde observations from more than 2,800 stations around the world~\citep{vandal2024global}. The evaluation on GDAS is calculated against 1000 stations with highest quality to avoid the impact of poor observations. Figure~\ref{fig:gdas_forecast}a shows the annual average RMSE skill for $z500$ and $t850$ variables over 10 days. All evaluations show the same trend, with FengWu-Adas having higher forecast RMSE than IFS-HRES in the early stage and lower than IFS-HRES starting around 7-8 days. Since Adas assimilates far fewer observations than the IFS, resulting in less accurate analysis, FengWu-Adas has a larger RMSE for forecasts at shorter lead times. However, FengWu accumulates errors more slowly for multi-step forecasts, so FengWu-Adas has a smaller RMSE for forecasts at longer lead times. Adas mainly provides the initial state for the forecasting model, which contributes more at shorter lead times, and FengWu contributes more at longer lead times. Figure~\ref{fig:gdas_forecast}b shows the box plot of the prediction RMSE with different lead times. Similar to the results of average RMSE in Figure~\ref{fig:gdas_forecast}a, FengWu-Adas performs slightly worse than IFS-HRES at shorter lead times, but performs better at longer lead times, with lower medians and IQRs. The scatter plot and corresponding fits between the observation and prediction values across the evaluation stations are illustrated in Figure~\ref{fig:forecast_regression}. The first two rows and last two rows correspond to the $z500$ and $t850$ variables, respectively. The columns have different lead times of 1, 5 and 10 days, and the coefficient of determination $R^2$ and the slope of the fitted line are marked in the subplot. The prediction of FengWu-Adas is more consistent with the observations at longer lead times and has a higher $R^2$. In addition, we visualize the distribution of forecast RMSE at the GDAS evaluation stations in Figure~\ref{fig:pred_rmse_station}. The initialization date-time is also selected as 2017-01-27 00:00 UTC, which is consistent with Figure~\ref{fig:visual_analysis}. The first, second and third columns correspond to the RMSE distributions with 1-day, 5-day and 10-day forecast lead times, respectively. It can be seen that in the forecast results with 10-day lead time, IFS-HRES has a large deviation in North America, while FengWu-Adas has a better forecast performance in this region. In general, FengWu-Adas shows accurate end-to-end global weather forecasts, demonstrating the promising potential of data-driven approaches.

\section*{Discussion}\label{Discussion}

Significant progress has been made in AI-based weather forecasting research on the ideal setting based on ERA5 dataset, which prompts our focus to shift towards end-to-end forecasting based on realistic observations. Some ambitious yet radical endeavors have aimed to discard the background and data assimilation components, striving for direct forecasting from observations as initial conditions~\citep{vaughan2024aardvark,mcnally2024data}. However, to date, the forecast-assimilation cycle remains the most reliable mainstream paradigm for operational end-to-end weather forecasting. This work primarily concentrates on data assimilation and end-to-end weather forecasting based on conventional observations (e.g., in-situ weather stations, ships and buoys, radiosonde observations), which serve as the most direct and accurate information sources. Nowadays, satellite observations are playing an increasingly vital role in data assimilation. Typically, they do not directly provide observations of geophysical variables but rather measurements such as radiation, brightness temperature, reflectivity, etc., which require additional inversion steps, although neural networks inherently suit this process~\citep{xu2024fuxi}. Compared with the pre-satellite era, the operational forecasting capabilities of meteorological centers using a complete global observing system including satellites have improved dramatically. Lacking the assimilation for satellite and radar observations, especially satellite data that are important for large-scale processes, constitutes a primary limitation of this study and a key direction for our future exploration. Another limitation lies in that our method is based on grid data of a given resolution, and interpolating the original off-grid data will inevitably introduce additional errors. Recently, techniques based on continuous spatial modeling have offered a promising train of thought to directly handle off-grid data and have achieved impressive results in scaling to arbitrary resolutions~\citep{chen2024fnp}. Moreover, this method of modeling implicit functional representation can support multi-modal data (e.g., in-situ observations and satellite data) to form a universal framework and can be readily extended to 4D data assimilation. In addition, since our method does not change the forecasting model, the end-to-end forecasts of the system still inevitably inherit its shortcomings, such as the blurred prediction. This problem is hopeful to be solved as the forecasting model improves.

In fact, ERA5 itself integrates various observational data, including conventional observations, satellites, and radars, in the process of being generated. Unlike the optimization process of traditional data assimilation methods, supervised learning makes this information inevitably part of the implicit priors introduced in model training. For the current end-to-end paradigm, the precision of the supervised labels determines the upper limit of the method's performance, posing a challenge to be addressed in the future. Producing high-quality analysis data and forecasts that may be superior to any existing product should be the ultimate goal of data assimilation and data-driven revolution, which still requires a lot of exploration and efforts. Recent study has shown promising progress and may become the mainstream paradigm for the next stage of the research in the domain of end-to-end weather forecasting~\citep{alexe2024graphdop}. The reanalysis datasets offer the highest-quality complete labels at present and hold an irreplaceable significance. In fact, high-quality data has always been essential for the development of AI and many other technologies. Therefore, the development of physical models and traditional algorithms remains crucial. The purpose of our study is not to replace the traditional NWP systems but to demonstrate the potential of AI methods in tackling real-world challenges. Both physical techniques and AI methods have their own strengths and limitations and should be treated equally and considered as powerful tools for benefiting science and society. We believe that this is a meaningful exploration and look forward to the deployment and implementation of AI methods in operational systems in the future.

\section*{Methods}\label{Methods}

\subsection*{Data preparation}\label{Data preparation}

ERA5 is used as the ground truth and source of the simulated observations in our experiments. ERA5 is a global atmospheric reanalysis archive containing hourly weather variables such as temperature, geopotential, wind speed, humidity, etc. A subset of the ERA5 dataset for 39 years, from 1979 to 2017, is chosen to train and evaluate the model. We choose to conduct experiments on a total of 69 variables at a resolution of 0.25° ($721\times1440$ grid points), including 5 upper-air variables with 13 pressure levels (i.e., 50hPa, 100hPa, 150hPa, 200hPa, 250hPa, 300hPa, 400hPa, 500hPa, 600hPa, 700hPa, 850hPa, 925hPa and 1000hPa), and 4 surface variables. Specifically, the upper-air variables are geopotential ($z$), temperature ($t$), specific humidity ($q$), zonal component of wind ($u$) and meridional component of wind ($v$), whose 13 sub-variables at different vertical level are presented by abbreviating their short name and pressure levels (e.g., $z500$ denotes the geopotential at a pressure level of 500 hPa), and the surface variables are 10-meter zonal component of wind ($u10$), 10-meter meridional component of wind ($v10$), 2-meter temperature ($t2m$) and mean sea level pressure ($msl$). We use the data from 1979-2015 for training, 2016 for validation, and 2017 for testing.

The real observational data are parsed from the prepbufr files of GDAS, which are archived by the National Centers for Environmental Information (NCEI). GDAS provides various types of observational data starting from 2012-02-13 00:00UTC, including multi-source data such as aircraft and upper-air reports, surface land/marine reports, scatterometer data, satellite-derived wind reports, etc. The prepbufr files consist of many messages with different types, each of which contains many observation columns with different height levels. The conventional types of observations in GDAS are used for data assimilation and evaluation in our experiments, covering the main geophysical variables such as temperature, humidity, and wind speed. Each observation contains the longitude and latitude coordinates, pressure level, time, and the value and quality marker of the geophysical variables. Before being input into the neural network, the observations are first processed into 0.25° grid data through nearest neighbor interpolation, and the grid values without observations are filled with 0. The confidence matrix is specified based on the quality marker, time offset, and interpolation distance to represent the quality of the observation value, and the confidence at places without observations is set to 0 to indicate filled spurious observation values. Regarding the large number of missing GDAS observations after 2018, data from 2017 are used for testing and others for training. 

\subsection*{Evaluation metrics}\label{Evaluation metrics}

The latitude-weighted root mean square error (RMSE) is a statistical metric widely used in geospatial analysis and atmospheric science, which evaluates the accuracy of a model's forecasts or estimates of variables across different latitudes. The latitude weighting is a common strategy to account for the varying area represented by different latitudes on a spherical Earth. The definition of RMSE is as follows.

Given the estimate $\hat{x}_{h,w,c}^t$ and its ground truth $x_{h,w,c}^t$ for the $c$-th channel at time $t$, the RMSE is defined as
\begin{equation}
\resizebox{0.7\hsize}{!}{$
\operatorname{RMSE}(c,t) = \sqrt{\frac{1}{H_0\cdot W_0}\sum\nolimits_{h,w} H_0 \cdot \frac{\operatorname{cos}(\alpha_{h,w})}{\sum_{h'=1}^{H_0} \operatorname{cos}(\alpha_{h',w})}(x_{h,w,c}^t - \hat{x}_{h,w,c}^t)^{2}}
$}
\end{equation}
where $h$ and $w$ denote the indices for each grid point along the longitudinal and latitudinal directions, respectively, and $\alpha_{h,w}$ is the latitude of point $(h,w)$. For the forecast $\hat{x}_{h,w,c}^{i+\tau}$ and its ground truth $x_{h,w,c}^{i+\tau}$ for the $c$-th channel with lead time $\tau$, the RMSE is defined as
\begin{equation}
\resizebox{0.8\hsize}{!}{$
\operatorname{RMSE}(c,\tau) =\frac{1}{T} \sum_{i=1}^{T} \sqrt{\frac{1}{H_0\cdot W_0}\sum\nolimits_{h,w} H_0 \cdot \frac{\operatorname{cos}(\alpha_{h,w})}{\sum_{h'=1}^{H_0} \operatorname{cos}(\alpha_{h',w})}(x_{h,w,c}^{i+\tau} - \hat{x}_{h,w,c}^{i+\tau})^{2}}
$}
\end{equation}
where $T$ is the total number of test time slots. The RMSE calculated at the stations is not latitude-weighted.

\subsection*{Deep network modules}\label{Deep network modules}

\paragraph{Patch embedding and patch recovery} 
The patch embedding and patch recovery are implemented through standard convolution and transposed convolution, respectively. The patch size of Adas in our experiments is set to 6 and the number of vertical layers remains unchanged. This means that a patch has $1\times6\times6$ pixels for 3D upper-air variables and $6\times6$ pixels for 2D surface variables. The kernel sizes and strides of 3D and 2D convolution are the same as the patch size, and the two parts are then concatenated together in the vertical direction. Therefore, in our experiments, $H=H_0/6$, $W=W_0/6$, $D=14$, and the dimension $C$ is 192. The parameters of patch embedding and patch recovery are not shared for each input.

\paragraph{Patch merging and patch expanding} 
In order to capture multi-scale meteorological features, the patch merging and patch expanding modules are used for down-sampling and up-sampling (Figure~\ref{fig:principle}d). Patch merging first rearranges pixels to reduce the horizontal dimensions by half, while the number of channels increases. After the regularization layer, the number of channels is then transformed by a linear layer. Therefore, patch merging achieves halving of the horizontal dimensions and doubling of the channel dimension. The patch expanding module performs the opposite operation and makes the encoder form a symmetric Unet~\citep{ronneberger2015u} structure. Similarly, the number of vertical layers is not changed in the process of patch merging and patch expanding.

\paragraph{Gated convolution} 
Gated convolution~\citep{yu2019free} is a convoluional operation based on soft-gating, which employs continuous masks ranging between 0 and 1 to represent the degree of validity for each pixel. Originally, the gating mask in gated convolution was learned automatically from the data at each layer, and the output features were weighted by it. 
In Adas, the confidence matrix of our method can serve as the gating mask. By using the confidence matrix as input for each layer and updating it explicitly, we propose an improved form of gated convolution to provide a more complete representation for sparse observations. The guidance of the confidence matrix can help the network extract features more effectively and capture the interactions between the background and observations more efficiently in the gated cross-attention module. Specifically, each value in the confidence matrix signifies the degree of confidence associated with corresponding observation values, and the confidence matrix is dynamically updated to guide the varying confidence levels of different layers. Let the subscript $i$ and $o$ represent the input and output of each layer, respectively, then the gated convolution and the update rule of the confidence matrix can be formulated as
\begin{equation}
    y_o = SiLU \big( Conv3d(y_i) \big) \odot Sigmoid \big( Conv3d(m_i) \big)
\end{equation}
\begin{equation}
    m_o = Sigmoid \big( Conv3d(m_i) \big)
\end{equation}
where $\odot$ represents element-wise multiplication. The $Sigmoid(\cdot)$ activation ensures that the confidence matrix values are between 0 and 1, and the activation function for the features can be arbitrary. We use the $SiLU(\cdot)$ activation here, which has been proven to be a better choice than $ReLU(\cdot)$~\citep{hendrycks2016gaussian}. The gated convolution module can provide a more complete representation for observations and improve the quality of feature fusion.

\paragraph{Gated cross-attention} 
The cross-attention mechanism is introduced to capture the interactions between the background and observations, but treating observations at different locations with different confidence levels equally is clearly unreasonable. Especially for observations with low confidence levels, directly calculating cross-attention may have negative guidance on the background. To address this issue, we propose the gated cross-attention module, which is also guided by the confidence matrix. Specifically, when the observations are used as the condition of cross-attention, the background is aligned with them at first, ensuring that observations influence the background in proportion to the confidence levels. After calculating the cross-attention, the remaining proportion of the background that does not participate in the operation will be added back. In contrast, the background fully influences observations since all values in the background generally have relatively high confidence levels, which can correct the observations with low confidence levels. Therefore, the confidence matrix needs to be updated accordingly, which is also implemented through a convolutional layer with $Sigmoid(\cdot)$ activation (Figure~\ref{fig:principle}c). 
The process of gated cross-attention can be expressed as 
\begin{equation}
    x^b_o = Attention(x^b_i \odot m_i, y_i, y_i) \oplus \big(x^b_i \odot (1-m_i)\big)
\end{equation}
\begin{equation}
    y_o = Attention(y_i, x^b_i, x^b_i)
\end{equation}
where the inputs of $Attention(\cdot)$ correspond to query ($Q$), key ($K$) and value ($V$), respectively, and $\oplus$ represents element-wise addition. The gated cross-attention module utilizes information for interactions selectively based on confidence levels, which can effectively avoid the negative impact of low-quality data.

\subsection*{Training loss and optimization}\label{Training loss and optimization}

The mean absolute error (MAE) loss, also known as L1 loss, is employed to supervise the training of the neural network. In multi-variate optimization, an issue of imbalanced optimization arises when there are significant differences in the magnitudes of losses for different variables. Therefore, it is necessary to weigh the losses for different variables. Instead of setting the weights manually, the losses for each variable (i.e., each channel) are automatically weighted to have the same magnitude in our training. The models in simulation experiments are trained for 50 epochs using the AdamW optimizer~\citep{loshchilov2018fixing} and OneCycleLR scheduler~\citep{smith2019super}. The learning rate starts from 1e-6, warms to a maximum value of 1e-4 for 10 epochs, and then decays gradually to 1e-10. To ensure adequate training with available GDAS observational data, the models in GDAS experiments are fine-tuned for 50 epochs based on the transfer learning strategy.

\subsection*{Inference speed}\label{Inference speed}

Compared to traditional methods, the reduction of computational consumption and execution time is a significant advantage of neural networks. In our system, Adas requires only about 0.6 seconds for the inference of data assimilation on an NVIDIA Tesla-A100 GPU, and FengWu needs about 1.5 seconds for inferring a single-step forecast. This means that our system can complete a periodic iteration in seconds once the observational data are available, which achieves significant efficiency improvement (about 1000-10000 times faster) compared to traditional data assimilation algorithms~\citep{zhang2019operational}.

\section*{Acknowledgements}

This research is supported by National Natural Science Foundation of China (No.62071127 and 62101137), National Key Research and Development Program of China (No.2022ZD0160101), Shanghai Natural Science Foundation (No.23ZR1402900), Shanghai Science and Technology Commission Explorer Program Project (24TS1401300) and Shanghai Municipal Science and Technology Major Project (No.2021SHZDZX0103). We acknowledge the ECMWF and NCEI for their great efforts in storing and providing the invaluable data, which are very important for this work and the research community. We would also like to express our appreciation to Dr. Ben Fei, Mr. Huihang Sun and Prof. Jiayuan Fan for their suggestions and valuable discussions during the conduct of this research, and the research team and service team in the Shanghai Artificial Intelligence Laboratory for the provision of computational resources and infrastructure.

\section*{Data availability}
The ERA5 dataset can be downloaded from the official website of Climate Data Store (CDS) at \url{https://cds.climate.copernicus.eu}. The GDAS observational bufr files can be obtained from the Archive Information Request System (AIRS) of NCEI at \url{https://www.ncei.noaa.gov/has/HAS.DsSelect}. The IGRA dataset and its documentation are publicly available at \url{https://www.ncei.noaa.gov/data/integrated-global-radiosonde-archive}.

\section*{Code availability}
The neural network model is developed using standard libraries in open-source platforms including PyTorch. Codes and model checkpoints used in this study are available at \url{https://github.com/OpenEarthLab/FengWu-Adas}.

\bibliographystyle{plainnat}  
\bibliography{references}  






\end{document}